# Prompt engineering for bibliographic web-scraping

Manuel Blázquez Ochando. manublaz@ucm.es Universidad Complutense de Madrid
https://orcid.org/0000-0002-4108-7531

Juan José Prieto Gutiérrez. jjpg@ucm.es Universidad Complutense de Madrid.
https://orcid.org/0000-0002-1730-8621

María Antonia Ovalle Perandones. maovalle@ucm Universidad Complutense de Madrid.
https://orcid.org/0000-0002-6149-4724

**Abstract**

Bibliographic catalogues store millions of data. The use of computer techniques such as web-scraping allows the extraction of data in an efficient and accurate manner. The recent emergence of ChatGPT is facilitating the development of suitable prompts that allow the configuration of scraping to identify and extract information from databases. The aim of this article is to define how to efficiently use prompts engineering to elaborate a suitable data entry model, able to generate in a single interaction with ChatGPT-4o, a fully functional web-scraper, programmed in PHP language, adapted to the case of bibliographic catalogues. As a demonstration example, the bibliographic catalogue of the National Library of Spain with a dataset of thousands of records is used. The findings present an effective model for developing web-scraping programs, assisted with AI and with the minimum possible interaction. The results obtained with the model indicate that the use of prompts with large language models (LLM) can improve the quality of scraping by understanding specific contexts and patterns, adapting to different formats and styles of presentation of bibliographic information.

Introduction

Large language models (LLM) have transformed the generation and understanding of natural language and have gained wide coverage and potential in the scientific community (Zhao et al., 2023). In this context of great advances, it is essential to know how to take advantage of the capacity of Artificial Intelligence (AI) to obtain the best results in the tasks entrusted to it. This is the design of prompts or, in other words, the textual input written by the user, with the instructions or questions posed in his or her query. The prompt sets the context of the conversation, tells the LLM model what information is important and what the desired output, form, transformations and content should be (Qi et al., 2023). Ambiguity, reinforcement of bias, overfitting, lack of context, ethical considerations are some of the key challenges that the information professional has to deal with in order not to get incomplete, incorrect, inaccurate or grossly misleading answers. This entails the generation of optimised, well-structured prompts that correctly represent the need, task, method or phases of work through which the AI action must pass. In fact, Gao et al. (2023) reveal that ChatGPT performance is highly dependent on the style and formal organisation of the message and semantic refinement (Zhu, X., Kuang, Z., & Zhang, L. 2023).

For this reason, prompts must be designed according to a method, which is still under development by the scientific community, and which some have dared to call 'prompt engineering', which was initially researched and popularised in LLMs by Liu et al. (2023). This justifies the need to develop a specific competence in prompt creation, leveraging the capabilities and features of LLMs, such as ChatGPT, facilitating more engaging and impactful interactions with these advanced language models. Such engineering actions have profound implications in software development (Khojah et al., 2024; Vaillant et al., 2024) supporting, as discussed above in software programming (Xia & Zhang, 2023). In this sense, prompt engineering involves creating and tuning specific instructions that guide the behaviour of the language model to obtain more accurate and consistent responses. This may include techniques such as zero-shot prompting, few-shot prompting, Chain-of-Thought (CoT) Prompting, Auto-CoT, Consistency and Coherence or Emotion Prompting, where specific examples are provided, so that the model learns and responds appropriately to new tasks.

Thus, as a Zero-shot prompting technique, the language model is asked to perform a task without providing specific input-output examples. The model uses its pre-existing knowledge acquired during its training to generate responses based on the given instruction. For example, if the model is asked to translate a sentence without having been given prior examples of translation, the model will attempt to use its understanding of the language and context to perform the task (Kojima et al., 2022; Kong et al., 2023). Few-shot prompting is when the model is provided with some input-output examples to help it better understand the task to be performed. These examples act as a guide for the model, improving its ability to generate accurate and relevant responses. For example, if you want the model to classify the sentiment of a sentence, you can give it a few examples sentences with their corresponding sentiment classification and then ask it to classify a new sentence (Reynolds & McDonell, 2021). Alongside these is Chain-of-Thought (CoT) prompting which is a technique that promotes step-by-step reasoning in language models.

Instead of generating a direct answer, the model is encouraged to decompose the problem into smaller steps and solve each step sequentially. An example of CoT oriented to our case could be the following: 'We need to extract bibliographic data from the catalogue of a university library. Answer by following these steps: 1) Identify the model of the catalogue card. 2) Identify the data container tags. 3) Design the web-scraping strategy. 4) Program the programme in PHP'. The AI model is provided with concrete steps, correlated and chained by logic, so that it has the steps that will allow it to solve a complex problem. This helps to make the answers more structured and easier to solve. For example, in the case of a mathematical problem, the model could detail each step of the calculation instead of just providing the final result (Wei et al., 2022). Auto-CoT is a method that automates the creation of chains of reasoning. This approach aims to improve model robustness and reduce errors by generating multiple distinct chains of reasoning. The most consistent or appropriate chain of reasoning is then selected. This automated process allows the model to explore various avenues of thought and improve the accuracy of its answers. An example of this can be found in the work of Huang et al. (2024) in which the extraction of events, with their temporal relationships and causal relationship, is chained into the input parameters of the LLM. In addition, Consistency and Coherence techniques focus on the 'Contrastive Chain-Of-Thought' to improve the consistency and coherence of the answers generated by the models. Consistency refers to the model's ability to maintain the same line of reasoning throughout a conversation or task, while coherence refers to the internal logic and flow of responses. These techniques help the model generate responses that are not only correct, but also cohesive and easy to understand (Nye et al., 2021). In the sphere of emotions, Emotion Prompting is a technique that manages not only emotions, but also tone in the model's responses, adapting them to different contexts and needs. This technique allows the model to respond in a more human and empathetic way, adjusting their tone according to the situation. For example, in a customer support conversation, the model may use a more understanding and calm tone if it detects that the user is frustrated or upset (Ul Huda et al., 2024).

Although there is previous work on AI script generation, such as that of Lázaro-Rodríguez (2024), our research differs in three fundamental aspects: a) Development of a structured prompt engineering method specifically for bibliographic web-scraping. b) Optimisation to obtain functional code in a single interaction. c) Extensive validation with a dataset of more than 55,000 records. d) Demonstration of the interoperability of the method between different AI models (ChatGPT-4o and Claude).

**Scope of study and objectives**

This paper presents an innovative method that revolutionises the way web-scrapers for bibliographic catalogues are developed, by applying advanced prompt engineering techniques with LLMs. The main objective is to achieve complete and efficient automation of functional code generation in a single interaction with ChatGPT-4o, eliminating the need for multiple iterations and manual refinements. The main objective will be achieved through the specific objectives listed below:

a) To achieve full automation in the coding of web-crawler programs oriented to Library and Information Science and Documentation, especially in the bibliographic context.
b) To design a query method for AI that significantly reduces the number of interactions required to obtain a fully functional web-crawler. It is generally known that a high number of corrections are required to achieve the software code that is being sought. This, in part, is due to poor interaction, which does not consider the correct definition of context, purpose and expected result.
c) Make bibliographic data-mining accessible with alternative methods to OAI-PMH, SRU-39.50. Researchers do not always have the most favourable conditions for data collection and web-scraping can be a tool at their disposal for this purpose.
d) To understand the behaviour, logic or reasoning of AI in the context of developing web-scraper programmes in a language such as PHP, resulting in an easy implementation on any web server.

e) To provide a method that allows the researcher to quickly customise and adapt the prompt for any altmetric and bibliometric research. In this way, the researcher is enabled in the achievement and development of their own data extraction tools.

**Methodology**

The method to achieve a correct development of a web-scraper programme in PHP with AI, necessarily involves the use of well-defined prompts.
The methodology is based on the combination of 'Role Prompting' and 'Few-shot Prompting', techniques selected for their proven effectiveness in code generation tasks (Yang et al., 2023; Nguyen et al., 2023). Role Prompting allows contextualising the expertise needed for the task, while Few-shot Prompting facilitates learning through specific examples from the bibliographic domain. This combination has been shown to be superior to other methods such as Zero-shot Prompting in specialised programming tasks (Kong et al., 2023). Our method seeks to minimise interactions with the AI by making the best possible use of the ChatGPT attention layer (Vaswani et al., 2017). It is important to consider that, as the conversation with the AI grows to better profile tasks, the attention layer may lose key details of the original task, leading to the risk of 'hallucination' and, consequently, unsatisfactory results (Huang et al., 2023; Duan, Yang & Tam, 2024; Yehuda et al., 2024; Verma et al., 2023). This forces the opening of new conversations and complicates the interrogation procedure, resulting in a waste of time and effort to achieve the initial goal. AI must be able to recognise the instructions and tasks necessary to create a programme, with good performance and even more, adapted to the particularities and needs of each case. To this end, some authors claim that it is possible to structure the input prompt or message through markdown (Atlas, 2023; Pividori & Greene, 2023; Greshake et al., 2023), in order to make the AI attention layer more effective in interpreting and executing tasks. The attentional layer is critical because it allows AI to focus on the relevant parts of the input, prioritising critical information and filtering out the superfluous. This differential attentional capability improves text comprehension and processing, allowing for a more accurate and relevant response to the prompt's instructions. As Vaswani et al. (2017) point out, attentional architecture is crucial for handling long-term dependencies in data streams, which is essential in the generation and understanding of complex instructions such as those required for web-scraper programming.
In this research we will work with the AI systems ChatGPT-4o and Claude Sonnet 3.5, following the next phases:

1) Selection of the web-scraping target. The catalogue datos.bne.es, which belongs to the National Library of Spain, has been selected due to the idiosyncrasy of the data it provides, i.e. the semantic scope, with structured and semi-structured access to the representation of the data on the web page. Moreover, according to the latest institutional report (BNE, 2021), this catalogue consists, in sum, of 17.4 million bibliographic records, authority entries, holdings and locations. This provides a suitable testing platform to demonstrate the importance of prompt design, and to improve AI performance in this environment. The method described can be applied to any bibliographic catalogue, regardless of country or content.

2) Definition of the control prompt, based on a direct query, to elaborate a web-scraper of the bibliographic catalogue designated as the target of analysis. A simple query, in the form of a direct sentence or phrase, is represented, which contains the program's generation order. This prompt will be used to compare the results obtained subsequently with those produced by the advanced prompt.

3) Design of the advanced prompt, according to the latest advances in the scientific literature on software development, in order to obtain the code base of the web-scraper in PHP applied to the indicated bibliographic catalogue. In this phase, the aim is to indicate in detail the task that the AI has to develop in order to create an optimal scraping programme for the case.

4) Quantitative and qualitative comparative analysis. The results are evaluated by means of:

- Success rate in generating functional code
- Number of interactions needed to obtain working code
- Accuracy in the extraction of bibliographic data.
- Execution time and performance of the generated code
- Comparative analysis of the code generated by both prompts (control and advanced) The evaluation is performed on a sample of 55,473 records from the National Library of Spain's catalogue, representing approximately 0.32% of the total available records.

5) Consolidation and verification of the web-scraper generated by the AI. This stage is dedicated to the execution of a stress test of the designed software, with the objective of capturing the first 50,000 records of the National Library's catalogue. This will allow the detection of any operational failures, for discussion and analysis. During the test, the code developed by ChatGPT-4o is executed sequentially, adapting the record identification variable automatically. As data is collected, it is recorded in a MySQL database, forming the test dataset. To facilitate the programming of the loop and the data insertion instructions that are required in addition, a specific prompt for the Claude Sonnet 3.5 AI will be used, which reaffirms the effectiveness of the prompt design method described here and its interoperability with other AIs. This approach ensures that any researcher can replicate and evaluate the work, ensuring the reliability and accuracy of the programme.

**Development**

The prompt, or AI input, is a critical phase in order to obtain a PHP web-scraper code base. Poor interaction with the AI can result in code that does not respond to the needs and technological support available. This usually happens in most cases when simple prompts are designed, based on direct questions, without planning or predefined structure (Sahoo et al., 2024; Chen, B.; Zhang, Z.; Langrené, N.; Zhu, S. 2023a). This leads to the misconception that AI produces more failures than it actually does. According to the method established for this research, a control prompt, representing a typical simple direct query, to generate a web-scraper for the case of the National Library of Spain's data portal datos.bne.es, would be as follows:

**Table 1.** Control prompt, representing a simple query to solve the problem of a web-scraper for datos.bne. https://github.com/manublaz/promptAI/blob/main/webScraping-biblioCatalog-test1-en.txt

> Create a web-scraping programme in PHP that is capable of extracting the bibliographic data from the following record on the data portal of the National Library of Spain: https://datos.bne.es/edicion/bimo0001291967.html

In this type of query, the user assumes many details that the AI does not know. Although the programming language to be used and the type of software to be developed are indicated, as well as the objective, not enough information is provided about the context, the method of data extraction and the particularities of the case of the bibliographic catalogue of the Spanish National Library. The AI has to make an exercise of imagination to determine that the PHP application context will be in a local, test, Apache type web server. It also has to infer that by referring to the "extraction of bibliographic data", the selection of typical bibliographic metadata such as title, author, edition, date of publication, publisher, collection, subject, among others, is intended. However, you are not being instructed to access the link to analyse the possible metadata used in the record of that record "bimo0001291967". Nor are you being told the possible description fields and their structure, so you can assume that they are represented either in table form, Dublin Core metadata, XML, MARC21, or any other format. By asking direct, too open-ended questions, it also has the effect of giving too much leeway to the AI. This means that it can choose the method it deems best to extract the required information from the catalogue record, which will not necessarily match the one the user requires. In this case, the response from the control prompt, available in table 2, shows that ChatGPT chose the "simple_html_dom.php" library. It did not make use of other common and recommended options such as "cURL", or "file_get_contents()". In this sense, there is a free will effect that depends directly on the knowledge base, the examples, the training and what the attention layer is able to understand at the control prompt. In the face of indefiniteness, AI tends to provide the combination of codes, best weighted for a general case of web-scraping.

**Table 2.** Response obtained by ChatGPT for the control prompt.
https://github.com/manublaz/promptAI/blob/main/webScraping-biblioCatalog-test1-en-results.php

```php
<?php
// Include the simple_html_dom library
require 'simple_html_dom.php';

// URL of the National Library of Spain record
$url = 'https://datos.bne.es/edicion/bimo0001291967.html';

// Create an instance of simple_html_dom
$html = file_get_html($url);

if(!$html) {
    die('Could not access the page');
}

// Extract the bibliographic data
$title = $html->find('h1[property="dc:title"]', 0);
$author = $html->find('span[property="dc:creator"]', 0);
$date = $html->find('span[property="dc:date"]', 0);
$publisher = $html->find('span[property="dc:publisher"]', 0);
$description = $html->find('div[property="dc:description"]', 0);

echo "Title: " . ($title ? $title->plaintext : 'Not available') . "\n";
echo "Author: " . ($author ? $author->plaintext : 'Not available') . "\n";
echo "Date: " . ($date ? $date->plaintext : 'Not available') . "\n";
echo "Publisher: " . ($publisher ? $publisher->plaintext : 'Not available') . "\n";
echo "Description: " . ($description ? $description->plaintext : 'Not available') . "\n";
?>
```

In this same table 2, we can see how the extraction of the bibliographic data is carried out on the basis of a "hypothetical" interpretation of the HTML source code of the record of the National Library of Spain. This is with <span> and <h1> tags with "property" attributes containing the prefix and name of the Dublin Core metadata for the bibliographic description. This parameter assumption is also an assumption based on the lack of information provided in the prompt. ChatGPT understands that the Dublin Core metadata context is typical of bibliographic catalogues in a national library context, but does not know exactly whether the actual source code of the record will actually be like this. Based on this, it proposes as a response an approximation that in no case adjusts to reality, generating in the user a perception of inaccuracy or lack of capabilities on the part of the AI. Thus, when executing the code provided as a result of the control prompt, it is not possible to obtain the data from the catalogue record, resulting in an "error" screen or unavailability of the contents.

There is consensus on the structure of the prompt needed to make ChatGPT capable of creating a test program or software (Brown et al., 2020; Yang et al., 2023; Nguyen et al., 2023). The following five processes should be highlighted: a) There should be a purpose and context section, clearly describing the problem, the development environment and the level of detail sought. b) Inputs and constraints should be expressed, to clearly determine what information the user will provide and what rules need to be followed in the processing of data or content. c) Examples and expected results, showing in a simple way what is expected from his work or from the instructions provided. d) Methodology, steps and procedures to carry out the orders or tasks given to the AI, indicating with precision and the necessary breakdown, what the programme code has to do, its workflow. e) Defining the type of output or output specifications of the programme or of the data to be prepared.

Taking into account these bases, the structure of the advanced prompt, for a web-scraper development case, is as shown in the following table 3:

**Table 3.** Advanced prompt used to generate the web-scraper for datos.bne.
https://github.com/manublaz/promptAI/blob/main/webScraping-biblioCatalog-test2-en.txt

---

**Role**
You are a researcher in the area of "Software Development" and "Documentation Sciences". Your expertise is in "web-scraping" and "PHP language", as well as in the functions "cURL" and "file_get_contents".

**Context and purpose**
# *Problem* Develop a web-scraper to extract data from the following web page:
https://datos.bne.es/edicion/bimo0001291967.html
# *Development environment* The web-scraper has to be programmed in "PHP language" and run in an "Apache environment", with "MySQL" database support.
# *Level of detail* I need the web-scraper program to allow the download of the HTML code of the web page, and extract the information from the tables it contains (title, place of publication, publisher, date of publication, physical description or extension, other physical characteristics, dimensions, type of material, symbol, location, headquarters), which it will store properly in a PHP array "$arrayPHP".

**Inputs and constraints**
# Use the following functions "curl_init", "curl_exec", "curl_setopt" when programming the web-scraper.
# Use "XPath" to extract data for title, place of publication, publisher, publication date, physical description or extent, other physical characteristics, dimensions, type of material, call number, location, and institution.
# You may need to use the function "preg_match" to check if the retrieved data row contains the metadata title we are looking for, where XXXX in <strong>XXXX</strong> is the name of the metadata that identifies the information we intend to extract (title, place of publication, publisher, publication date, physical description or extent, other physical characteristics, dimensions, type of material, call number, location, institution).

**Input and output examples**

# Input example
<tr><td class="label-row"><strong>Título</strong></td> <td>El "profundo Isaac" ;documentos inéditos del archivo de Isaac Peral y Caballero ;recopilación de hechos y documentos efectuada por su hijo Antonio ;</td></tr>
 <tr><td class="label-row"><strong>Lugar de publicación</strong></td><td>Madrid</td></tr> ...

# Output example

$title="El profundo Isaac | documentos inéditos del archivo de Isaac Peral y Caballero | recopilación de hechos y documentos efectuada por su hijo Antonio"; $publishPlace="Madrid";

echo " <li>$title - $lugarPublicacion</li> ";

$arrayPHP[]=array("title"=>'El profundo Isaac | documentos inéditos del archivo de Isaac Peral y Caballero | recopilación de hechos y documentos efectuada por su hijo Antonio', "publishPlace"=>'Madrid');

**Detailed steps**

> 1. Process the $url="https://datos.bne.es/edicion/bimo0001291967.html";
> 2. Use cURL to obtain the HTML code.
> 3. Apply XPath to obtain the data rows from the table.
> 4. Use preg_match to discriminate the metadata and add the information to the arrayPHP.
> 5. Print the results on the screen.

In the first instance, the role of the AI is indicated, both from the point of view of its performance and its specific skills and knowledge. The idea behind this section is that you select the contents related to these keywords for the development of your tasks. In this case "Software Development" is specified, to express that you are a programmer specialised in coding "web-scraping" with the specific desired language "PHP".

Familiarity with the 'cURL' and 'file_get_contents' functions is explicitly specified for specific technical reasons:

cURL offers critical advantages for bibliographic web-scraping:

- Greater control over HTTP connections
- Advanced cookie and session management
- Ability to handle redirects and SSL certificates
- Better error handling and timeouts
- Support for authentication and custom headers

file_get_contents, although simpler, is mentioned because:

- It is a common alternative in basic scraping scripts.
- Allows the AI to understand the level of complexity required
- It facilitates comparison with more robust methods
- It helps direct the AI towards the most appropriate solution.

The deliberate mention of both functions in the prompt serves as implicit technical guidance, directing the AI towards the use of cURL, which is more appropriate for the systematic extraction of large-scale bibliographic data. This preference is subsequently reinforced in the specific constraints of the prompt.

In the "context and purpose" section, we seek to define three key aspects, namely: a) Problem, b) Development environment, c) Level of detail. Firstly, the "problem" that the AI has to solve. In this case, the development of a web-scraper to obtain the data of a specific web page. The URL of the target website is not mandatory, but it can provide some of the context it needs, as it can recognise the domain name and the extension or format of the content. For example, by processing "*datos.bne*" it can deduce that it is a catalogue of the National Library of Spain, which means that the scraper's application context is focused on a catalogue with bibliographic description metadata. In the "development environment" also known as the working environment, the context of execution of the program is made explicit, in this case, an Apache HTTP server, PHP compiler, and MySQL, i.e., the classic distribution of any web server. This further restricts the framework for developing the program. With regard to the "level of detail", the level of detail desired for the extraction, the metadata to be obtained from the HTML source code obtained from the target web page, as well as the storage to be provided for the data are indicated. At this stage of development, you choose to record the data in the form of an array.

In the section "Inputs and restrictions", the restrictions or aspects that you will have to comply with in the development of the code are indicated. For example, you are obliged to use the functions "curl_init", "curl_exec", "curl_setopt" so that the extraction method is entirely cURL and not "file_get_contents". The restriction works, because even though the role also knows about this alternative method, it is explicitly told to use cURL. In fact, in the result obtained, it discards "file_get_contents" in favour of cURL. The XPath constraint has also been added, as the preferred method for selecting the nodes containing the desired metadata, which are listed according to the order in which they appear on the target web page. Finally, an "optional" constraint is added to freely determine whether or not its use is appropriate for the achievement of the goal. This is the "preg_match" function used to check or evaluate regular expression patterns. Since the data are not identified by attributes or classes in the HTML code, it is required to obtain the rows of the table in which they are found and detect the presence of the keyword in one of the cells, which contains the name of the metadata between the tags <strong>XXXX</strong>, where XXXX is the location of the metadata. In this way, we create a recognition pattern in the AI, where XXXX must be replaced by one of the metadata presented to it (title, place of publication, publisher, date of publication, physical description or extent, other physical characteristics, dimensions, type of material, symbol, location, venue). We could call this technique "unknown replacement technique" and consider it an essential operation within broader chains of reasoning in AI. As can be seen, this technique consists of replacing placeholders with specific values, and although in itself a simple operation, it can be part of more complex inference processes, where multiple logical steps and deductions are made to reach a final goal.

Nevertheless, it is indispensable to provide "Input and output examples" so that you can interpret for what kind of code you will have to design the web-scraper. That is why an input example is provided, which should contain a sample or snippet of the HTML source code of the target web page, with the data and metadata in question. As can be seen, the <strong> tag appears and can make the association or inference that this is the row with the desired content. Next, the output example is provided, for that same input, creating a variable with the name of the metadata, which contains the desired data, a screen printout of the same and the aforementioned array or list, in which the information of the entire record is stored. To conclude the prompt, a summary is made of the working method of the code, also known as the "Detailed Steps" section, in which the phases that the program flow has to go through are listed in greater or lesser detail. The result is shown in Table 4 below.

**Table 4.** Response obtained by ChatGPT for the advanced prompt.
https://github.com/manublaz/promptAI/blob/main/webScraping-biblioCatalog-test2-en-results.php

```php
<?php
// URL of the webpage to scrape
$url = "https://datos.bne.es/edicion/bimo0001291967.html";

// Function to obtain the HTML content of the page using cURL
function getHtmlContent($url) {
    $ch = curl_init();
    curl_setopt($ch, CURLOPT_URL, $url);
    curl_setopt($ch, CURLOPT_RETURNTRANSFER, true);
    curl_setopt($ch, CURLOPT_FOLLOWLOCATION, true);
    $html = curl_exec($ch);
    curl_close($ch);
    return $html;
}

// Obtain the HTML content of the page
$html = getHtmlContent($url);

// Create a new DOMDocument object
$dom = new DOMDocument();
@$dom->loadHTML($html);

// Create a new DOMXPath object
$xpath = new DOMXPath($dom);

// XPath expressions to select the table rows
$rows = $xpath->query("//tr");

// Initialise the array to store the data
$data = [];

// Function to clean and normalise the text
function cleanText($text) {
    return trim(preg_replace('/\s+/', ' ', $text));
}

// Iterate through the rows and extract the data using preg_match
foreach ($rows as $row) {
    $label = $xpath->query("td[@class='label-row']/strong", $row)->item(0);
    $value = $xpath->query("td[2]", $row)->item(0);

    if ($label && $value) {
        $labelText = cleanText($label->nodeValue);
        $valueText = cleanText($value->nodeValue);

        switch ($labelText) {
            case 'Título':
                $data['title'] = $valueText;
                break;
            case 'Lugar de publicación':
                $data['placeOfPublication'] = $valueText;
                break;
            case 'Editorial':
                $data['publisher'] = $valueText;
```

```php
            break;
        case 'Fecha de publicación':
            $data['publicationDate'] = $valueText;
            break;
        case 'Descripción física o extensión':
            $data['physicalDescription'] = $valueText;
            break;
        case 'Otras características físicas':
            $data['otherPhysicalCharacteristics'] = $valueText;
            break;
        case 'Dimensiones':
            $data['dimensions'] = $valueText;
            break;
        case 'Tipo de material':
            $data['materialType'] = $valueText;
            break;
        case 'Signatura':
            $data['signature'] = $valueText;
            break;
        case 'Localización':
            $data['location'] = $valueText;
            break;
        case 'Sede':
            $data['headquarters'] = $valueText;
            break;
      }
    }
}

// Print the results on screen
foreach ($data as $key => $value) {
    echo "<li><strong>$key:</strong> $value</li>";
}

// Print the array to verify the data
echo '<pre>';
print_r($data);
echo '</pre>';
?>
```

As can be seen, ChatGPT creates a function to get the HTML code of the target web page, applying all the restrictions indicated, regarding the functions it has to use. It then makes use of the DOM function and with XPath obtains the path to the rows containing the metadata and data in question. It even takes the liberty to consider a text cleaning and normalisation function, which is quite correct, for the purpose of data processing (Chai, 2023). Much more interesting, however, is the development of the foreach loop in which it analyses the rows it has been able to extract from the data table, where it selects with XPath the metadata tag and its associated data in the adjoining cell. However, for the purpose of differentiating text from metadata tags, it does not use "preg_match", it uses the "switch() {case}" function to distinguish the different cases. This is well within the prompt, as he was given the choice to use it or not, and as we can see, he has decided that it was a better method, we should understand, because of its simplicity of application. This allows the hypothesis and reflection on the autonomy of the AI, when making the optimal decisions in the scenarios that arise, which is not without concern, since, by repeating the operation, but adding the constraint of obligatoriness, in effect, it ends up using the "preg_match" function. However, the first version provided fully satisfies the requirements and needs expressed.

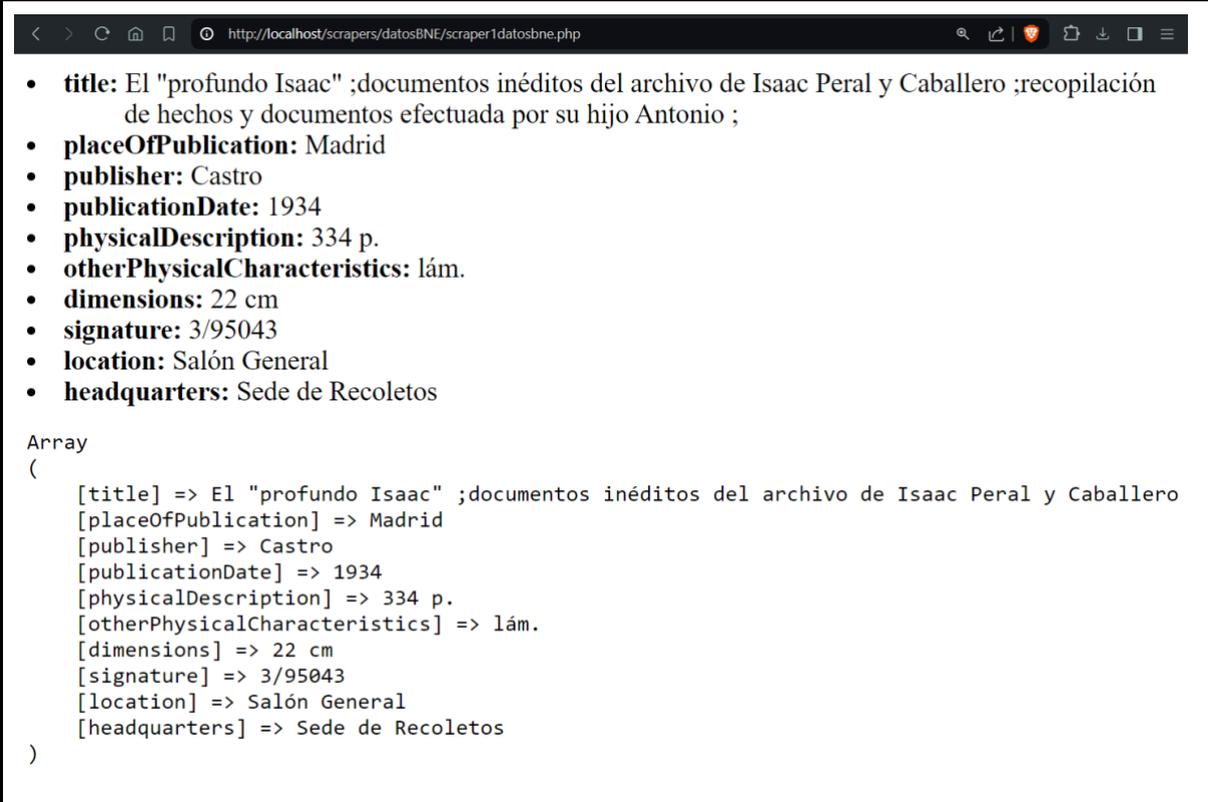

**Fig. 1.** Result obtained from executing the code provided by ChatGPT in the advanced prompt.

Finally, the code expresses the screen printout of each metadata with its data, and the generated array, which has been given the name "$data" and not "$arrayPHP", which is not a significant error. It can be summarised, therefore, as a satisfactory experience, very correct and without errors, as can be seen in the execution of the code in figure 1.

Next, the stress tests were carried out, which consolidate and verify the functioning of the web-scraper obtained with ChatGPT. The objective was to collect the first 50,000 records of the data catalogue in order to detect possible errors, omissions or malfunctioning of the code. However, in order to be able to run the web-scraping program in a loop, iterating each of the catalogue record identifiers, the following changes are required: (a) add codes for database connection setup, create and verify connection; (b) introduce a function to handle null or empty values, in case the information source does not provide data; (c) create a main loop to loop through the BNE data catalogue records; (d) add the insert SQL code and run it for a table structure similar to the data collected by the web-scraper; (e) add a 3-second stop after each insert, so as not to cause a concurrency problem on the server. To address these changes efficiently and ensure the correct functioning of the prompting method, we have reused its original structure, and added the changes and requirements to be made on the web-scraping program. This time, unlike the previous one, the prompt will be processed by a different artificial intelligence, specifically Claude Sonnet 3.5. This change is crucial to assess the capability of the prompting method in terms of interoperability, i.e. its ability to be interpreted correctly regardless of the AI model used. The adjusted prompt is shown in table 5.

**Table 5.** Prompt executed in Claude Sonnet 3.5, to add the looping function and insert records into the database, reusing the web-scraping program generated by ChatGPT. https://github.com/manublaz/promptAI/blob/main/webScraping-biblioCatalog-test3-es.txt

---

**Rol**
You are a researcher in the area of 'Software Development' and 'Documentation Sciences'. Your expertise is specialised in 'web-scraping' and 'PHP language'.

**Context and purpose**
# *Problem* To reuse the web-scraping program provided, adding the necessary PHP codes, in a way that allows its execution in a loop to collect the bibliographic records of the catalogue.
# *Development environment* The software is programmed in 'PHP language' and runs in an 'Apache environment', with 'MySQL' database support.

**Entries and restrictions**.
- Respect the code provided.
- Add code for database connection setup, create and verify connection.
- Create a main loop to loop through the BNE data catalogue records.
- Create a variable $number to format the counter variable of the main loop, with leading zeros, to compose a 10-digit number with leading zeros, thus https://datos.bne.es/edicion/bimo$number.html using the 'str_pad' function.
- Introduce a function to handle null or empty values, in case the data source does not provide data.
- The MySQL table is called 'datosBNE' and its structure is the following 'url, author, title, placeOfPublication, publisher, publicationDate, physicalDescription, otherPhysicalCharacteristics, dimensions, materialType, signature, location, headquarters'.
- Add the SQL insertion code and execute it for the indicated table structure, loading the information collected in the web-scraping process.
- Add a stop of 3 seconds after each insertion, so as not to cause a concurrency problem on the server.
- In case the value of the $data['title'] is unspecified or null, then jump to the next url in the loop.

**Detailed steps**
1. Add database connection to the code.
2. Create a main loop to loop through the catalogue records.
3. Manage the preparation of the SQL query and its execution to insert records into MySQL.

---

As can be seen, the main structure of 'role', 'context and purpose', 'inputs and constraints' and 'detailed steps' is maintained, with the deletion of the 'Input and output examples' section. In this case, the most relevant are the limits and restrictions that set the pattern and conditions that must be met. First of all, 'respect the code provided', as together with this prompt, the file with the web-scraping programme developed by ChatGPT in PHP language is attached. The aim is for the AI to respect this code and use it to elaborate the loop that allows iterating the library catalogue records. On the other hand, the changes noted in the previous paragraph are stated, relating to the connection with the database, the control of the loop counter, which modifies the $url variable, with which the successive records of the catalogue are revised, as well as its particular case of being a ten-digit numbering. In this case, note that the AI is forced to use the 'str_pad' function, indicating with double quotes the key texts or data for the attention layer. On the other hand, since the program has to handle cases where empty fields are encountered, it is indicated that it must consider a method or function to handle null values. This is important because it is a chained reasoning with the key condition to register records

in the database, i.e. that the main field obtained in the variable '$data[title]' is different from an empty or null value. This chaining helps the AI to relate both guidelines and consider that the handling of null values determines the registration or continuation of the loop with the next record in the catalogue. It is also provided with details of the database table, as well as its structure, so that it can generate the corresponding SQL insert query. Accessorily, although it would not be necessary to indicate it, this SQL construction is reinforced by indicating that the information collected by the web-scraping program should be inserted and executed in the table. In this way, the AI already infers that it is the collected data that matches the name of the variables and fields in the database table. This prompt was executed in the Claude Sonnet 3.5 AI, to test whether the method of constructing prompts, described in this paper, is interoperable.

The result could not have been more satisfactory. Table 6 shows the source code returned by Claude. As can be seen, it respects the guideline of programming the connection with the database, for which it creates the connection variables, and a PDO object that allows the connection with the database to be executed. On the other hand, it does respect the web-scraping code made by ChatGPT, which shows that it correctly assimilates what this implies and assumes its implementation in a new code with the additions that are outlined. It is also interesting, the method of handling null or empty values, which integrates correctly, considering the array '$data[]' of data storage of the scraping program. On the other hand, it respects the instructions given about the pauses, and the condition to insert records in the database, depending on the null or empty value of the record title. The whole set is correctly integrated in a loop with a maximum limit of ten digits.

**Table 6.** Program resulting from using the ChatGPT code and the one provided by Claude 3.5
https://github.com/manublaz/promptAI/blob/main/webScraping-biblioCatalog-test3-es-results.php

```php
<?php
// Database connection configuration
$host = 'localhost';
$dbname = 'DB_name';
$username = 'your_DB_username';
$password = 'your_DB_password';

// Create connection
try {
    $pdo = new PDO("mysql:host=$host;dbname=$dbname", $username, $password);
    $pdo->setAttribute(PDO::ATTR_ERRMODE, PDO::ERRMODE_EXCEPTION);
    echo "Successful connection to the database.<br>";
} catch(PDOException $e) {
    die("Connection error: " . $e->getMessage());
}

// Function to get the HTML content of the page using cURL
function getHtmlContent($url) {
    $ch = curl_init();
    curl_setopt($ch, CURLOPT_URL, $url);
    curl_setopt($ch, CURLOPT_RETURNTRANSFER, true);
    curl_setopt($ch, CURLOPT_FOLLOWLOCATION, true);
    $html = curl_exec($ch);
    curl_close($ch);
    return $html;
}

// Function to clean and normalize text
function cleanText($text) {
    return trim(preg_replace('/\s+/', ' ', $text));
}

// Function to handle null or empty values
function handleNullValue($value) {
    return ($value !== null && $value !== '') ? $value : null;
}

// Prepare the SQL insertion query
$sql = "INSERT INTO datosBNE (url, author, title, placeOfPublication, publisher, publicationDate, physicalDescription, otherPhysicalCharacteristics, dimensions, materialType, signature, location, headquarters) VALUES (?, ?, ?, ?, ?, ?, ?, ?, ?, ?, ?, ?, ?)";
$stmt = $pdo->prepare($sql);

// Main loop to iterate over records
for ($i = 1; $i <= 9999999999; $i++) {
    $number = str_pad($i, 10, '0', STR_PAD_LEFT);
    $url = "https://datos.bne.es/edicion/bimo{$number}.html";

    // Get the HTML content of the page
    $html = getHtmlContent($url);

    // Create a new DOMDocument object
    $dom = new DOMDocument();
    @$dom->loadHTML($html);
```

```php
// Create a new DOMXPath object
$xpath = new DOMXPath($dom);

// XPath expressions to select table rows
$rows = $xpath->query("//tr");

// Initialize the array to store data
$data = [
    'url' => $url,
    'author' => null,
    'title' => null,
    'placeOfPublication' => null,
    'publisher' => null,
    'publicationDate' => null,
    'physicalDescription' => null,
    'otherPhysicalCharacteristics' => null,
    'dimensions' => null,
    'materialType' => null,
    'signature' => null,
    'location' => null,
    'headquarters' => null
];

// Loop through the rows and extract data
foreach ($rows as $row) {
    $label = $xpath->query("td[@class='label-row']/strong", $row)->item(0);
    $value = $xpath->query("td[2]", $row)->item(0);

    if ($label && $value) {
        $labelText = cleanText($label->nodeValue);
        $valueText = cleanText($value->nodeValue);

        switch ($labelText) {
            case 'Título - Title':
                $data['title'] = $valueText;
                break;
            case 'Lugar de publicación - Place of publication':
                $data['placeOfPublication'] = $valueText;
                break;
            case 'Editorial - Publisher':
                $data['publisher'] = $valueText;
                break;
            case 'Fecha de publicación - Publication date':
                $data['publicationDate'] = $valueText;
                break;
            case 'Descripción física o extensión - Physical description':
                $data['physicalDescription'] = $valueText;
                break;
            case 'Otras características físicas - Other physical characteristics':
                $data['otherPhysicalCharacteristics'] = $valueText;
                break;
            case 'Dimensiones - Dimensions':
                $data['dimensions'] = $valueText;
                break;
            case 'Tipo de material - Material type':
                $data['materialType'] = $valueText;
                break;
            case 'Signatura - Signature':
                $data['signature'] = $valueText;
                break;
            case 'Localización - Location':
                $data['location'] = $valueText;
                break;
```

```php
        case 'Sede - Headquarter':
            $data['headquarters'] = $valueText;
            break;
      }
    }
  }

  // If the title is null or not specified, skip to the next URL
  if (handleNullValue($data['title']) === null) {
    continue;
  }

  // Prepare values for insertion, handling null values
  $values = array_map('handleNullValue', array_values($data));

  // Execute the SQL query
  try {
    $stmt->execute($values);
    echo "Record inserted successfully: {$url}<br>";
  } catch(PDOException $e) {
    echo "Error inserting record: " . $e->getMessage() . "<br>";
  }

  // 3-second pause
  sleep(3);
}

// Close the connection
$pdo = null;
?>
```

Considering this code, a test was carried out to verify the correct functioning of the programme under real operating conditions. To this end, a data compilation target was established in the datosBNE catalogue, until 50,000 bibliographic records were reached. The result of this test can be consulted in the dataset
https://github.com/manublaz/datasets/blob/master/dataset-datosbne-50k_2024-07-19.sql
The analysis of the performance of the web-scraping programme applied to the bibliographic catalogue of the National Library of Spain (datosBNE) reveals significant aspects of its efficiency, considering the restrictions implemented to preserve the integrity of the server. The programme in table 6 examined 62,786 links, of which 7,313 (11.65%) resulted in 404 errors. A total of 55,473 records were collected over a total period of 63 hours, 3 minutes and 12 seconds, including scheduled pauses of 3 seconds between each insertion. The effective scraping time, excluding these pauses, was approximately 13.85 hours. This substantial difference between the total time and the effective time underlines the impact of the mitigation measures implemented to protect the server. The effective extraction rate is calculated at 66.76 records per minute or 4,005.60 records per hour. The figures compiled in table 7 represent the actual performance of the scraper, discounting the time of scheduled breaks.

**Table 7.** Performance results obtained by the web-scraping programme generated by ChatGPT and Claude. SQL queries used to perform the calculations presented here. https://github.com/manublaz/datasets/blob/master/dataset-datosbne-50k_2024-07-19_SQLqueries-es.txt

| Metric | Value |
| --- | --- |
| Total number of links reviewed | 62.786 |
| 404 links from theBNE data catalogue | 7.313 |
| Total records collected | 55.473 |
| Start date | 2024-07-16 15:47:57 |
| End date | 2024-07-19 11:16:57 |
| Total duration | 67:29:00 |
| Number of catalogue service anomalies, not related to the programme | 1 |
| Duration of the anomaly | 4:25:48 |
| Total duration discounting the anomaly | (63 hours, 3 minutes and 12 seconds) = 227.592 seconds |
| Scheduled breaks | 3 seconds per record |
| Total time of scheduled breaks | 46:13:39 |
| Effective scraping time | (16 hours, 49 minutes and 33 seconds) = 60,573 seconds |
| Theoretical ideal time to execute the scraping process | 55,473 records * 3 seconds = 166,419 seconds |

A total of 5,407 jumps in record numbering were detected, with an average of one jump every 2,3521 records. The range of these jumps varies from 2 to 81 records, with a standard deviation of 2.5433, suggesting considerable variability in the structure of the BNE database. Additional waiting times between insertions, beyond the scheduled pause, were relatively rare. Most operations were completed within the expected time, with only 14 cases requiring more than 20 seconds, probably due to jumps in catalogue record numberings. The average completion rate per record of 85.4301% reflects the level of detail of the information available in the BNE data. This value suggests that, on average, the records contain a substantial amount of information, although there is still room for improvement in data completeness.

The comparative analysis of the web-scraper efficiencies reveals a significant contrast between its effective performance (calculated as the quotient between the ideal theoretical time and the effective scraping time 166,419 / 60,573 seconds) of 274.74% and its real performance (calculated in a similar way, but using the total time without anomalies 166,419 / 227,592 seconds) of 73.12%. This disparity quantifies the impact of the mitigation measures implemented, mainly the 3-second scheduled pauses between insertions. The high effective efficiency demonstrates the scraper's ability to process data quickly, exceeding 2.7 times the expected baseline speed. On the other hand, the actual efficiency of 73.12% reflects the trade-off between performance and server protection, indicating that the total execution time is longer than the theoretical ideal due to the intentional pauses, but still allows for meaningful and sustainable data collection.

**Limitations**

This research has some limitations that should be considered:

- Domain specificity: The method has been validated mainly with the catalogue of the National Library of Spain, although the principles should be applicable to other bibliographic catalogues.

- Technology dependency: The generated code is limited to PHP and the specified technology stack (Apache, MySQL).

- Evolution of LLMs: The results are linked to the specific versions of the models used (ChatGPT-4o and Claude).

- Performance constraints: The 3-second scheduled pauses, although necessary to respect the servers, have an impact on the total execution time.

**Discussion**

The findings of this research have important practical implications that significantly transform the work of researchers, librarians and documentation centres. In the field of bibliometric and documentary research, researchers have traditionally faced significant obstacles in collecting large bibliographic datasets (Robinson-Garcia, et al, 2017). The methodology developed eliminates the need for advanced programming skills, allowing researchers without deep technical expertise to develop their own data mining tools. For example, a researcher studying the evolution of artificial intelligence publications in the National Library catalogue can easily adapt the generated code to extract and analyse thousands of records in a matter of hours, a task that traditionally required weeks of software development or expensive commercial tools.

For librarians and documentalists, the impact is equally significant on their daily operations. Data migration between systems, a critical but often problematic task, is considerably simplified (Dula & Ye, 2012). An illustrative case is that of university libraries that need to synchronise their catalogues with national or international repositories. The proposed method allows for the rapid development of customised scripts for this task, reducing errors and processing time. In addition, it facilitates the identification of inconsistencies in bibliographic records; for example, during our tests with the BNE catalogue, the system automatically identifies variations in the way authors and titles are recorded, allowing a more efficient standardisation of data.

Library institutions particularly benefit in situations where access to specialised APIs or protocols such as OAI-PMH is unavailable or limited. During our research, we observed that many small and medium-sized libraries lack the resources to implement commercial data exchange solutions. Our method provides a viable and cost-effective alternative; the generated code can be easily adapted to work with different catalogue systems, as we demonstrated by successfully modifying the original script for the BNE.

In the development of document services, the implications are especially promising for innovation in library services. For example, a university library could use an adapted version of our system to create an automated bibliographic alert service. It would be able to monitor new acquisitions in various national catalogues on a daily basis, notifying researchers of relevant publications in their areas of interest. This type of value-added service was previously unfeasible due to technical and resource constraints.

Practical implementation of this methodology requires only a basic web server with PHP, a minimum technical requirement that most institutions already have. During our tests, the average implementation time was less than one day, including code adaptation. This technical accessibility, combined with the ability to generate functional code in a single interaction with the AI, represents a significant advance in the democratisation of advanced document management tools.

The quantitative results of our study support these practical implications: the extraction rate of 66.76 records per minute and the average completion rate of 85.43% demonstrate the real feasibility of the system for large-scale practical applications. These numbers translate into the ability to process entire medium-sized catalogues in a matter of days, a substantial improvement over traditional methods of bibliographic data extraction.

Automated extraction of bibliographic data through web scraping requires careful consideration of ethical and legal issues (Krotov, et al, 2020). In this research, we have implemented specific safeguards to ensure responsible use of library resources. The 3-second scheduled pause between requests is not arbitrary; it represents an ethical commitment to avoid overloading the servers of library institutions. In addition, our method respects the fair use policies of bibliographic catalogues, limiting the rate of extraction and avoiding interfering with other users' normal access. It is essential that researchers and practitioners implementing this methodology consider three basic principles: 1) consult and respect the data usage policies of each institution, 2) implement appropriate access rate control mechanisms, and 3) use the extracted data exclusively for research or library development purposes. Transparency in the extraction process, documenting the methodology and purpose of data collection, is essential to maintain trust between library institutions and the research community.

**Conclusions**

This work presents an effective working method to develop web-scraping programs, assisted with AI and with the minimum possible interaction. It is shown that a correctly configured prompt allows to obtain a debugged code, from the first interaction, satisfying the needs of data extraction and processing, starting from a given web resource, in this case, from a bibliographic catalogue. The AI is able to interpret the instructions and understand the subtleties of the task, tracing a workflow, assisted by a well-defined methodology. This demonstrates the importance of defining prompts with a clear method and structures, recognisable to the AI, that are likely to capture its 'attention' on the key data and procedures of the work it will have to develop.

The prompt is the indispensable element in the interaction with the AI. In the context of software development and more specifically of web-scraping programs, it is recommended to use the sections: a) Role, b) Context and purpose, c) Input and constraints, d) Input and output examples, e) Detailed steps. This confirms the observations made by the scientific community (Brown et al., 2020; Yang et al., 2023; Nguyen et al., 2023).

Constraints are essential to get the IA to work on the lines specified by its operator. It has been observed that all guidelines are adhered to, and where optional guidelines are provided, the AI can determine whether or not to consider them. This indicates that the results it provides are driven by the given rules and by other principles, probably that of minimum effort, simplicity or simplicity in the development of the code. In the specific case presented here, the 'switch' function was chosen instead of 'preg_match', to avoid the use of regular expressions and the excess of conditional structures that this would have entailed. Therefore, in the absence of further confirmation or experimentation, it could be that the AI knows rules of efficiency and effectiveness in code design, which could be a very useful feature for all researchers. This means that it is not necessary to be a specialist in programming to create programs that satisfy specific or specific needs, typical of scientific development, and in this case, of text and data mining.

Another aspect that proves fundamental is the input and output section of the prompt design. The AI is able to associate the details given in the context and in the tasks or instructions it has to perform, in the source code provided to it. Therefore, if it is shown an example of how it has to deal with the information or data, the chances of it providing a correct solution will be higher. In theory, it is a training provided to you, so that you can solve the specific problem of web-scraping code adapted to a very particular situation. While there is no consensus as to the number of input and output examples, it is understood that more input and output examples will enhance learning. However, the user has to be aware that not all AIs may have a sufficient context window to retain all details of all examples (Chen, S.; Wong, S.; Chen, L.; Tian, Y. 2023b; Dong et al., 2024). Therefore, it is suggested to include enough for a good understanding, probably between one and two. In this case, one example was sufficient. There is also the aspect of the variety of casuistries or exceptions presented in the examples, which often make AI training difficult. In such cases, it seems advisable to consider them in the methodology or context instructions of the prompt.

The methodological overview section of the prompt is crucial, as it ensures the continuous presence of the context, instructions and purpose of the task, from start to finish (Giray, 2023;

Ye et al., 2023). This ensures that the attentional layer of the AI retains the most important fundamentals, procedures and flows throughout the entire interaction. A well-structured methodological overview not only provides a clear and coherent framework for the task at hand, but also makes it easier for the AI to maintain a constant focus on the essential elements, avoiding deviations or loss of critical information. In designing prompts for AI systems, it is critical to consider how information is presented and how focus is maintained throughout the process. By including a methodological overview, a kind of 'anchor' is created that guides the model's attention to the key points that need to be addressed. This is especially important in complex or multi-stage tasks, where clarity and consistency are vital for successful implementation. In addition, this methodological approach helps mitigate potential confusion or errors that could arise if the AI loses sight of an important component of the task. By continuously reinforcing context and instructions, it promotes greater accuracy and efficiency in response.

The AI-generated web-scraper has proven to be a robust and efficient tool for bibliographic data extraction. It managed to process 55,473 records with an efficiency of 73.12% over the ideal theoretical time, efficiently handling 7,313 404 errors and 5,407 skips in record numbering. The extraction rate of 66.76 records per minute and the average completion rate per record of 85.4301% underline a good performance. The balance achieved between processing speed and respect for server limitations highlights that the code generated by ChatGPT and Claude has a good balance between simplicity, performance and effectiveness in data extraction. These results underline the potential of AI solutions in the development of web-scraping tools for large bibliographic datasets, capable of operating optimally within the technical and ethical constraints indicated by the researcher.